%% file: paper.tex
\PassOptionsToPackage{dvipsnames}{xcolor}
\PassOptionsToPackage{table}{xcolor}

\documentclass[sigconf]{acmart}

\usepackage{enumitem}
\usepackage{amsmath}
\usepackage{multirow}
\usepackage{booktabs}
\usepackage{scalerel}
\usepackage[linesnumbered,ruled,vlined]{algorithm2e}

\definecolor{textgray}{gray}{0.65}

\newcommand{\name}{iTrace}

\def\markup{1}
\if\markup1

\else

\fi

\AtBeginDocument{%
  \providecommand\BibTeX{{%
    \normalfont B\kern-0.5em{\scshape i\kern-0.25em b}\kern-0.8em\TeX}}}

\acmConference[GI '25]{Graphics Interface 2025}{May 28--30, 2025}{Toronto, Canada}
\acmBooktitle{Proceedings of Graphics Interface 2025 (GI '25), May 28--30, 2025, Toronto, Canada}
\acmYear{2025}
\copyrightyear{2025}
\acmISBN{978-1-4503-XXXX-X/25/05} 
\acmDOI{10.1145/XXXXXXX.XXXXXXX}  

\begin{document}

\title{\name\ : Interactive Tracing of Cross-View Data Relationships}


\author{Abdul Rahman Shaikh}
\email{ashaikh2@niu.edu}
\orcid{0000-0002-6046-4638}
\affiliation{%
 \institution{Northern Illinois University}
 \city{DeKalb}
 \state{Illinois}
 \country{USA}
}

\author{Maoyuan Sun}
\email{smaoyuan@niu.edu}
\orcid{0000-0002-0990-2620}
\affiliation{%
 \institution{Northern Illinois University}
 \city{DeKalb}
 \state{Illinois}
 \country{USA}
}

\author{Xingchen Liu}
\email{xliu7@niu.edu}
\affiliation{%
 \institution{Northern Illinois University}
 \city{DeKalb}
 \state{Illinois}
 \country{USA}
}

\author{Hamed Alhoori}
\email{alhoori@niu.edu}
\affiliation{%
 \institution{Northern Illinois University}
 \city{DeKalb}
 \state{Illinois}
 \country{USA}
}
\author{Jian Zhao}
\email{jianzhao@uwaterloo.ca}
\orcid{0000-0001-5008-4319}
\affiliation{%
 \institution{University of Waterloo}
 \city{Waterloo}
 \state{Ontario}
 \country{Canada}
}

\author{David Koop}
\email{dakoop@niu.edu}
\affiliation{%
 \institution{Northern Illinois University}
 \city{DeKalb}
 \state{Illinois}
 \country{USA}
}

\begin{abstract}
Exploring data relations across multiple views has been a common task in many domains such as bioinformatics, cybersecurity, and healthcare.
To support this, various techniques (e.g., visual links and brushing \& linking) are used to show related visual elements across views via lines and highlights.
However, understanding the relations using these techniques, when many related elements are scattered, can be difficult due to spatial distance and complexity.
To address this, we present \name, an interactive visualization technique to effectively trace cross-view data relationships.
\name\ leverages the concept of \textit{interactive focus transitions}, which allows users to see and directly manipulate their focus as they navigate between views. 
By directing the user's attention through smooth transitions between related elements, \name\ makes it easier to follow data relationships. 
We demonstrate the effectiveness of \name\ with a user study, and 
we conclude with a discussion of how \name\ can be broadly used to enhance data exploration in various types of visualizations.
\end{abstract}



\begin{teaserfigure}
  \includegraphics[width=\textwidth]{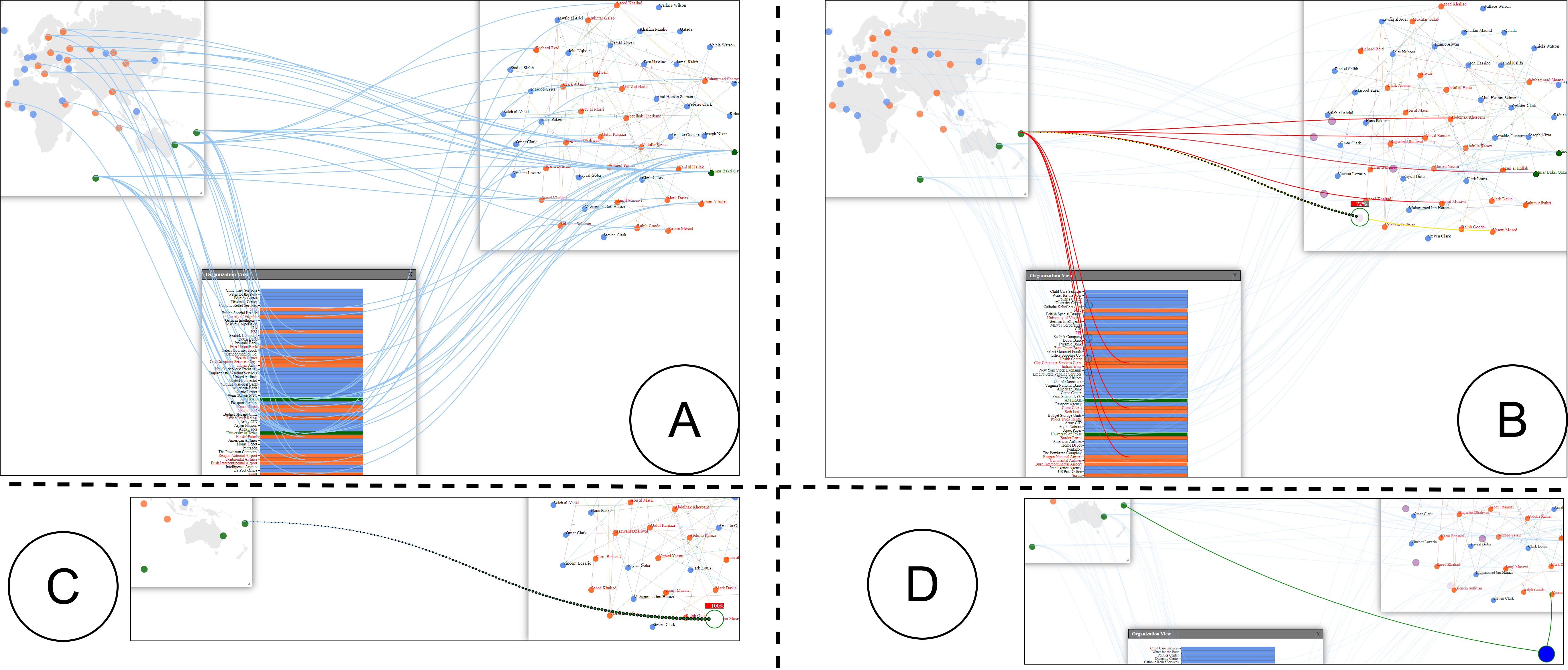}
  \caption{ An example scenario using \name: (A) an initial stage before tracing, (B) tracing an element from a map to a graph, where supportive foci (circles on red lines) automatically follow their visual links, and dynamic transitions adjust other links to reduce clutter, (C) the final stage where the user finishes tracing, and (D) manual link management that to organize links.
  }
  \Description{}
  \label{fig:teaser}
\end{teaserfigure}


\maketitle

\input{tex/1Intro}

\input{tex/2RelatedWorks}

\input{tex/3Design}
\input{tex/4Technique}
\input{tex/5Study}
\input{tex/6Discussion}

\begin{acks}
This work was supported in part by the NSF Grant IIS-2002082.
\end{acks}

\bibliographystyle{ACM-Reference-Format}
\bibliography{paper.bib}

%
%
%
%
%
%
%
%

\end{document}

%% file: tex/1Intro.tex
\begin{figure*}[tb]
  \centering
  \includegraphics[width=\textwidth]{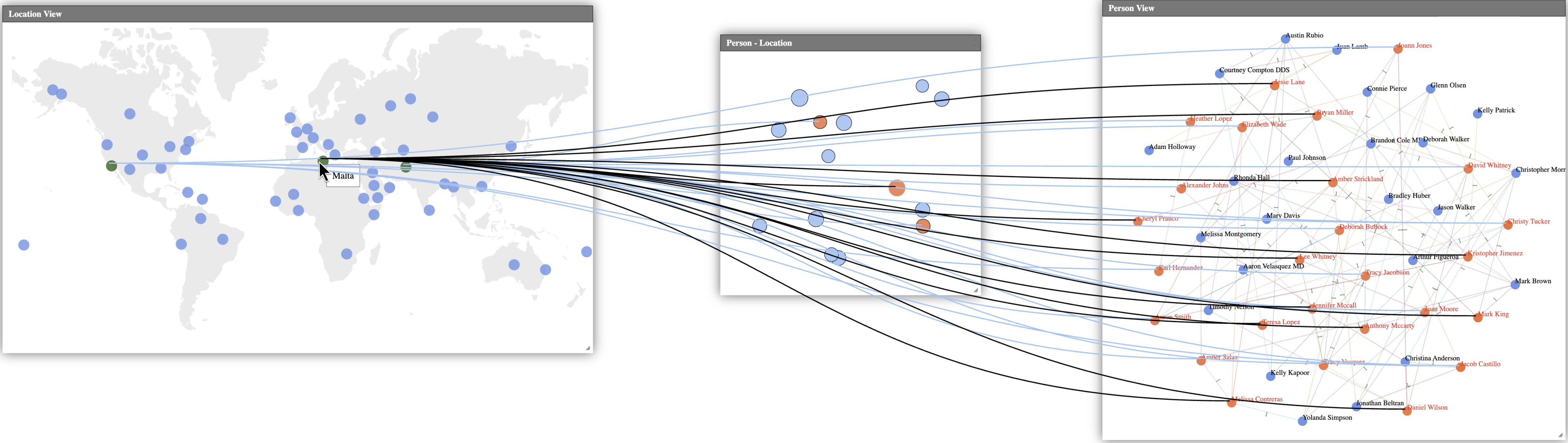}
    \vspace{-7mm}
  \caption{
  An example of the challenge on exploring related visual elements across views with visual links and highlights.
  }
  ~\label{challenge}
  \vspace{-1mm}
\end{figure*}

\section{Introduction}

\textit{Multiple-view visualization} (MV) is widely used for visual analysis in many fields (e.g., bioinformatics \cite{lex2010caleydo,gratzl2014domino, kale2023state}, cybersecurity \cite{chen2018user, zhang2015visualizing}, finance \cite{chang2007wirevis, jeong2008evaluating}, health care \cite{du2016eventaction}, 
and education \cite{zhao2018flexible, chen2017data}). 
To gain a deep understanding of the data, analysts often need to explore data relations across views \cite{sun2021towards, sun2021sightbi}. 
For example, cybersecurity analysts use network graphs to see device communications, lists to view suspicious organizations, and maps to check geographical locations, requiring them to explore relations across these views.
To support such explorations, techniques like visual links and brushing \& linking are heavily used.
Visual links connect related elements across views, and brushing \& linking highlight related elements after selections.
However, the spatial distribution of related visual elements can make it hard to follow links or interpret highlights, especially when many connections overlap or when elements are scattered.
Figure \ref{challenge} shows an example of the challenge, which explores related visual elements in a graph from a selected location on a map.

Previous attempts to address this have used aggregation-based techniques to reduce the clutter of visual links \cite{sun2021towards, sun2021sightbi, prouzeau2019visual, sun2019interactive, sun2018effect, sun2015biset, steinberger2011context}.
Although decreasing the number of visual links reduces overlap, it does not eliminate the need for users to transition across views, as the layout of views remains unchanged. 
As users shift across views to explore cross-view relationships, they still rely on visual links. 
While fewer intersections make it easier to trace a single link, users must pay close attention to avoid losing track of the link. 
Any lapse in focus can disrupt their ability to realign with the original pathway. 
This challenge is amplified when tracing multiple links simultaneously. 
Without visual guidance, it is hard to navigate complex cross-view relationships effectively, even with fewer visual links.
Techniques based on brushing \& linking eliminate the need for visual links, thereby avoiding issues with line crossings.
However, they struggle to explicitly outline the complex relationships between elements across views, especially after multiple selections. 
As each selection turns a group of elements in MVs into highlights, multiple brushings result in highlighting various groups. 
Relying only on highlights without visual links makes it difficult for users to discern relationships among elements across views. 


To address these challenges, we introduce \name, an interactive visualization technique to facilitate tracing cross-view data relationships. 
Unlike conventional methods that leave users to piece together disjointed visual cues, \name\ features the design of \textit{interactive focus transition}.
This mechanism allows users to visibly transfer their “focus” from one view to another via an on-screen marker, allowing them to see, manipulate, and track their focus while navigating across views. 
Continuous visual guidance ensures that even when multiple links must be traced simultaneously, the active pathway remains clearly highlighted, preventing users from losing track of connections.
Moreover, the versatility of \name\ makes it applicable to a wide range of visualizations, extending its usefulness beyond traditional multi-view scenarios.
In summary, our contributions are threefold. 
First, we conduct an in-depth design analysis to support tracing cross-view data relationships, expanding the design space of MVs. 
Second, we introduce a novel design concept, the \textit{interactive focus transition}, which emphasizes the visualization and direct manipulation of user focus points during cross-view data exploration. 
Third, we develop a visualization prototype, \name, which embodies our design, and we evaluate its effectiveness through a user study.

%% file: tex/2RelatedWorks.tex
\section{Related Work}

\begin{figure}[tb]
  \centering
  \includegraphics[width=\columnwidth]{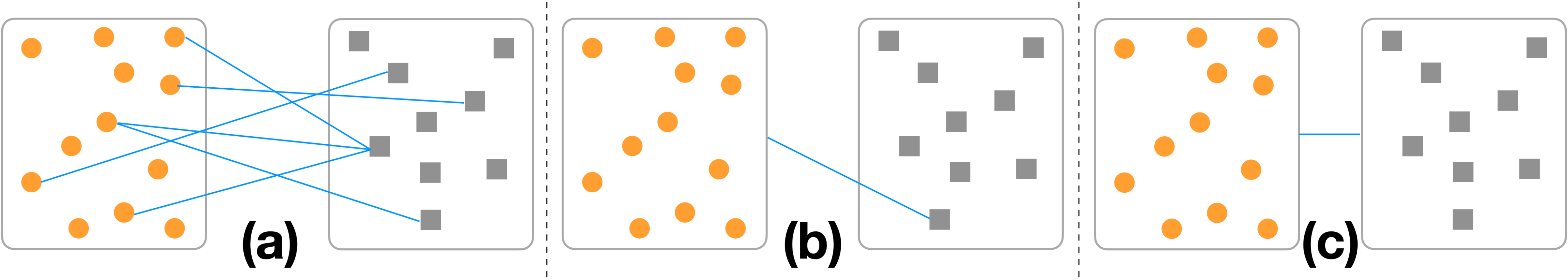}
    \vspace{-5mm}
  \caption{3 types of cross-view data relationships: (a) \textit{between visual elements}, (b) \textit{between visual elements and views}, and (c) \textit{between views}. 
 Each box shows a view, orange circles and gray squares are visual elements, and blue lines indicate relations.
  }
  ~\label{3links}
  \vspace{-2mm}
\end{figure}

\subsection{Cross-View Data Relationship}
\label{cvrelation}

Cross-view data relationships can be grouped into three categories (Figure \ref{3links}). 
The first involves relationships \textit{between visual elements} across views, focusing on the detailed connections between elements across views.
It examines interactions within individual views and how elements from separate views are connected. 
For example, it explores how five people in a social network graph might be linked to six organizations in another view.
This approach effectively ``breaks'' the boundaries of individual views to highlight components of potential knowledge graphs \cite{ji2021survey}, linking disparate data elements scattered across views. 
The second category covers relationships \textit{between visual elements and views},  offering a more abstract connection. Here, views add context or detail to visual elements in other views. For example, a line chart might show housing sales trends over the past six months for regions highlighted on a map. 
This relationship suggests a hierarchical structure among views, aligning with Shneiderman's visual information-seeking mantra \cite{shneiderman1996eyes}, where one view gives an overview and others provide details as needed. 
The third category looks at broader connections \textit{between views}, prioritizing overall insights from collective data rather than individual elements. For example, documents placed together may indicate relevance, showing patterns that span multiple components. This approach highlights more conceptual relationships for deeper understanding \cite{north2006toward}.

\begin{table}[tb]
\caption{Four levels of cross-view data relationships \cite{sun2021sightbi}.}
  \vspace{-2mm}
\label{4levels}
\resizebox{\columnwidth}{!}{
\begin{tabular}{r|c|c}
\multicolumn{1}{c|}{\textbf{Relationship Level}} & \textbf{Number of Views} & \textbf{Number of Visual Elements} \\ \hline
\textit{Individual} level $(1 : 1)$            & 2          & 2               \\ \hline
\textit{Group} level $(1 : m)$                 & 2          & $1 + m\ (m > 1)$             \\ \hline
\textit{Bi-group} level $(m : n)$              & 2          & $m + n\ (m, n > 1)$           \\ \hline
\textit{Multi-group} level $(m : n : ... : k)$ & $> 2$ & $m + n + ... + k \ (m, n, ..., k > 1)$ 
\end{tabular}
}
\end{table}

This work focuses on those \textit{between visual elements}, which require investigations of visual elements inside views. 
These low-level details across views form a key foundation for extracting higher-level insights, which are crucial to a data-driven sensemaking process \cite{pirolli2005sensemaking}.
The relationships can be categorized into four levels (Table \ref{4levels}), based on the cardinality of involved sets of visual elements from different views \cite{sun2021sightbi}.
The complexity increases from the \textit{individual} level to the \textit{multi-group} level \cite{sun2021towards, sun2014five}.

\subsection{Visualizing Data Relationships across Views}

Three designs for visualizing cross-view data relationships are \cite{sun2021towards}: \textit{connection}, \textit{visual highlight}, and \textit{spatial proximity}.

\textbf{Connection} uses visual links to show cross-view data relationships.
It draws lines among related visual elements across different views, corresponding to a one-to-one cross-view data relationship.
Examples include VisLink \cite{collins2007vislink} and its variant versions \cite{steinberger2011context, geymayer2014show}. 
It is widely used in visual analysis tools (e.g., Bixplorer \cite{fiaux2013bixplorer}, MyBrush \cite{koytek2017mybrush}, and Flowstrates \cite{boyandin2011flowstrates}). 
However, visual links can lead to clutter, making it hard for users to trace connections and understand more complex structures. 
To mitigate this, some studies have introduced aggregation techniques that group visual links into clusters (e.g., edge bundles) based on specific rules \cite{sun2021towards, sun2021sightbi, prouzeau2019visual, sun2015biset, steinberger2011context}.
While these clusters simplify the presentation, users still need to trace individual lines to fully understand the underlying relationships. 
Notably, previous works have also addressed navigation and focus management across visual links. 
For instance, Moscovich’s link sliding and bring \& go approaches \cite{moscovichlink} as well as Baudisch’s Drag-and-Pop/Drag-and-Pick technique \cite{baudischdrag} and the CompaRing method \cite{bertiniComparing} similarly explore the concept of drawing copies of data entities to pull them toward the user’s focus.

\textbf{Visual highlight} relies on the coordination of MVs, with a dynamic update strategy to show relationships between visual elements across views.
When users interact with elements in one view, corresponding elements in other views are highlighted. 
It is used in coordinated MVs \cite{roberts2007state, boukhelifa2003coordination, north1997taxonomy}, and is applied in various tools (e.g., Cross-Filtered Views \cite{weaver2009cross}, Improvise \cite{weaver2004building}, Interver \cite{sun2016interver}, MissBiN \cite{zhao2019missbin}, and Jigsaw \cite{stasko2007jigsaw}).
While this technique helps users identify related elements, it requires continuous attention to visual updates triggered by user interactions.
If these updates are missed, understanding the relationships can be hard. 
Unlike the connection approach, visual highlight avoids clutter by avoiding extra visual markers like lines, but without explicit links makes it hard to check complex cross-view relations.

\textbf{Spatial proximity} uses spatial distance to reveal relationships across views. 
It follows the ``near equals similar" visual metaphor, which depends on the spatial arrangement of views to present their data relationships.
This design requires users to interpret the spatialization of views to understand cross-view data relationships. 
It has been used in tools (e.g., Bixplorer \cite{sun2014role}, ForceSpire \cite{endert2012semantic}, GraphTrail \cite{dunne2012graphtrail},
and NodeTrix \cite{henry2007nodetrix}), which support ``space to think" \cite{andrews2010space} oriented sensemaking activities. 
As spatial arrangement is controlled at the view level (i.e., manipulating the layout of MVs \cite{shaikh2022toward}) instead of the visual element level, it is hard for users to identify and understand cross-view data relationships in detail.
Moreover, interpreting spatializations requires more cognitive effort than tracking lines or visual highlights. 


\subsection{Visualizations for Tracing Information}

\begin{table}[tb]
\caption{A Summary of Visualizations for Tracing}
\label{table-tracing}
\vspace{-2mm}
\resizebox{\columnwidth}{!}{%
\begin{tabular}{cc|lll}
\cline{3-5}
\multicolumn{2}{l|}{} &
	\multicolumn{3}{c}{\cellcolor[HTML]{FFCE93}\begin{tabular}[c]{@{}c@{}}\textbf{Tracing Target}\\ (number of involved \textit{visual element})\end{tabular}} \\ \cline{3-5} 
\multicolumn{2}{l|}{\multirow{-2}{*}{}} & \multicolumn{1}{c|}{\textbf{One Element}} & \multicolumn{2}{c}{\textbf{Multiple Elements}} \\ \hline
\multicolumn{1}{c|}{\cellcolor[HTML]{DAE8FC}} &
	\textbf{Data} &
	\multicolumn{1}{c|}{One entity} &
	\multicolumn{1}{c|}{One group} &
	\multicolumn{1}{c}{Multiple group} \\ \cline{2-5} 
	\multicolumn{1}{c|}{\cellcolor[HTML]{DAE8FC}} & 
	\textbf{Relation} & 
	\multicolumn{1}{l|}{\begin{tabular}[c]{@{}l@{}}- LineUp\cite{gratzl2013lineup}\\- Parallel Tag\\Clouds \cite{collins2009parallel} \\- Scatterplot\\Matrix \cite{carr1987scatterplot}\end{tabular}} & 
	\multicolumn{1}{l|}{\begin{tabular}[c]{@{}l@{}}- BubbleSets \cite{collins2009bubble} \\- Crossets \cite{perin2014manipulating} \\- Kelpfusion\cite{meulemans2013kelpfusion} \\- LineSets \cite{alper2011design} \\- Scatterplot\\Matrix \cite{carr1987scatterplot}\end{tabular}} &    
	\begin{tabular}[c]{@{}l@{}} - BiDot \cite{zhao2017bidots}\\
    - ChemoGraph \cite{kale2023chemograph}\\- Mercer \cite{wu2018interactive}\\- MyBrush \cite{koytek2017mybrush}\\- Sankey diagram \cite{riehmann2005interactive}\\- VisLink\cite{collins2007vislink} \end{tabular}   \\ \cline{2-5} 

\multicolumn{1}{c|}{\cellcolor[HTML]{DAE8FC}} & 
	\begin{tabular}[c]{@{}c@{}}\textbf{Location}\\(spatial\\transition)\end{tabular} & 
	\multicolumn{1}{l|}{\begin{tabular}[c]{@{}l@{}}- SoccerStories \cite{perin2013soccerstories}\\- Traffic flow\\visualizations \cite{andrienko2017visual}\\- Visual \\sedimentation \cite{huron2013visual}\end{tabular}}  & 
	\multicolumn{1}{l|}{\begin{tabular}[c]{@{}l@{}}- SoccerStories \cite{perin2013soccerstories}\\- Traffic flow\\visualizations \cite{andrienko2017visual} \\- Trajectory\\bundling\cite{du2015trajectory}\end{tabular}} &         
	\begin{tabular}[c]{@{}l@{}}- Andromeda\cite{self2018observation}\\-Animated \\transitions \cite{heer2007animated}\\- Traffic flow\\visualizations \cite{andrienko2017visual} \end{tabular} \\ \cline{2-5}
	
\multicolumn{1}{c|}{\multirow{-13}{*}{\cellcolor[HTML]{DAE8FC}\begin{tabular}[c]{@{}c@{}}\textbf{Analysis}\\ \textbf{Focus}\\(analytical\\space)\end{tabular}}} &
	\begin{tabular}[c]{@{}c@{}}\textbf{Time}\\(temporal\\trend, or\\ process)\end{tabular} &
 	\multicolumn{1}{l|}{\begin{tabular}[c]{@{}l@{}}- \`A Table \cite{perin2014table}\\- DimpVis \cite{kondo2014dimpvis}\\- egoSlider \cite{wu2015egoslider} \\- OpinionFlow \cite{wu2014opinionflow}\\- StoryFlow \cite{liu2013storyflow}\\- ThemeRiver \cite{havre2002themeriver} \end{tabular} } &
 	\multicolumn{1}{l|}{\begin{tabular}[c]{@{}l@{}}- EgoLines \cite{zhao2016egocentric}\\- EventThread \cite{guo2017eventthread}\\- Matrixwave \cite{zhao2015matrixwave}\\- Story Curves \cite{kim2017visualizing}\\- VisTrails \cite{callahan2006vistrails}\end{tabular}} &
  	\begin{tabular}[c]{@{}l@{}}- EgoLines \cite{zhao2016egocentric}\\- EventThread \cite{guo2017eventthread}\\- Reducing snapshots \\ to points \cite{van2015reducing}\\- Matrixwave \cite{zhao2015matrixwave}\\- Story Curves \cite{kim2017visualizing}\end{tabular} \\ 
\end{tabular}%
}
\end{table}

Visualization techniques for tracing information focus on 4 aspects of visual analysis: 1) \textit{data}, 2) \textit{relation}, 3) \textit{location}, and 4) \textit{time} (Table \ref{table-tracing}).
These mainly use \textit{visual links}, \textit{visual highlights}, and \textit{animation}. 
Tracing involves individual or multiple visual elements that encode data from one or more groups.
Such elements serve as tracing targets.
Their encoded data corresponds to the analysis focus in the data space.
This aligns with the information foraging loop in sensemaking \cite{pirolli2005sensemaking}.  
The tracing-oriented analysis explores relationships, spatial transitions, and temporal trends 
based on the encoded data. 

\textbf{Relation-focused tracing} involves analyzing connections between data entities
(e.g., one-to-one, one-to-many, many-to-many).
Simple cases involve tracing single visual elements \cite{collins2009parallel, carr1987scatterplot, gratzl2013lineup} across different visual contexts (e.g., lists, scatterplots, small multiples \cite{van2013small}), using visual links or highlights.
More complex cases trace multiple elements to identify those within the same group or across multiple groups.
For the former, set visualization techniques \cite{collins2009bubble, meulemans2013kelpfusion, alper2011design} use lines or ribbons to connect elements of the same set, requiring users to follow them to trace visual elements of the same set.
Brushing \& linking based highlights are also used for tracing groups of elements in different visual contexts (e.g., scatterplot matrix \cite{carr1987scatterplot}).
For the latter, visual links require users to follow multiple lines to trace elements across groups \cite{kale2023chemograph, wu2018interactive, koytek2017mybrush}.
However, such techniques lack effective guidance, expecting users to manually trace paths. 

\textbf{Location-focused tracing} 
examines spatial transitions by analyzing the movement of visual elements (e.g., sports or urban traffic analysis \cite{perin2013soccerstories, andrienko2017visual}).
Two main approaches are used: 1) line-based trajectories \cite{perin2013soccerstories, andrienko2017visual, du2015trajectory} where lines show paths, 
and 2) animations \cite{huron2013visual, du2015trajectory, self2018observation} 
showing dynamic transitions.
Line-based trajectories allow users to follow movement patterns, 
and multiple lines can be bundled by spatial proximity.
Animations guide user attention to moving elements, helping them stay focused. 
However, tracing animated elements is hard without visible trajectories, especially when elements move in different directions.
Additionally, most techniques lack effective animation controls, forcing users to switch between traced elements and interface controls (e.g., play and pause), interrupting the tracing process and complicating task resumption. 

\textbf{Time-focused tracing} involves temporal trends and process-oriented analysis, navigating through time to understand data evolution.
Temporal trends are shown with lines \cite{perin2014table, kondo2014dimpvis, wu2015egoslider} or areas \cite{wu2014opinionflow, havre2002themeriver}, where users follow shapes to observe value changes. 
For process-oriented analysis, stories, provenance, or event sequences are visualized using lines \cite{liu2013storyflow, zhao2016egocentric} 
, node-link diagrams \cite{callahan2006vistrails, van2015reducing}, and connected matrices \cite{zhao2015matrixwave}. 
Users follow line segments or paths to understand these processes.
By adjusting spatial layouts (e.g., ordering lines \cite{zhao2016egocentric} or matrices \cite{zhao2015matrixwave}), similar processes are placed near each other for comparison. 
However, users must manually trace paths without supportive features, which is hard with multiple processes.

In summary, while existing techniques use visual links, highlights, and animations, they lack assistance features and struggle to scale with multiple elements, especially for cross-view relationships. This motivates our design of \name.

%% file: tex/3Design.tex
\section{Designing \name\ }

\subsection{Term Clarification: Visual Link and Tracing}
\label{def-link-trace}

Our notion of \textbf{visual link} refers to a \textit{perceptually continuous geometric shape} (e.g., a solid, dashed, or dotted line, curve, or ribbon) that connects related data scattered across different locations on a screen. 
These links are visual marks displayed over existing representations to explicitly connect related elements.
This aligns with the definition of visual link discussed in \cite{steinberger2011context}, which highlights using additional visual marks instead of manipulating specific visual channels of existing visual marks (i.e., brushing and linking). 

We consider \textbf{tracing} as an \textit{analysis task} that guides users through conceptual spaces to find necessary information for sensemaking.
For example, users can follow colored contours in a scatterplot to explore a \textit{data space} and identify clusters, track edges in a graph to navigate a \textit{relationship space} or observe line segments in a chart to examine a \textit{time space} for trends. 
Specifically, in the context of cross-view data relations, tracing involves \textit{following visual encodings to identify connections between visual elements across views}, which helps users understand relationships between visual elements.

\subsection{User Tasks and Types of Tracing}
In the context of cross-view data relationships, there are three major types of user tasks \cite{sun2021sightbi}: \textit{filtering}-oriented, \textit{refocusing}-oriented, and \textit{connecting}-oriented tasks.
\textit{Filtering}-oriented tasks refer to using visual elements in one view to filter those in other views.
\textit{Refocusing}-oriented tasks involve exploring the same data across different views based on selections made in one view. 
\textit{Connecting}-oriented tasks aim to find connections between visual elements across views.
These tasks correspond to data filtering, data identification, and data connection, respectively.
As they all involve visual elements across views, tracing related elements serves as a key scaffold supporting the tasks. 
They are performed based on cross-view data relationships, particularly those between visual elements, as outlined in Section \ref{cvrelation}.


\begin{figure}[tb]
  \centering
  \includegraphics[width=\columnwidth]{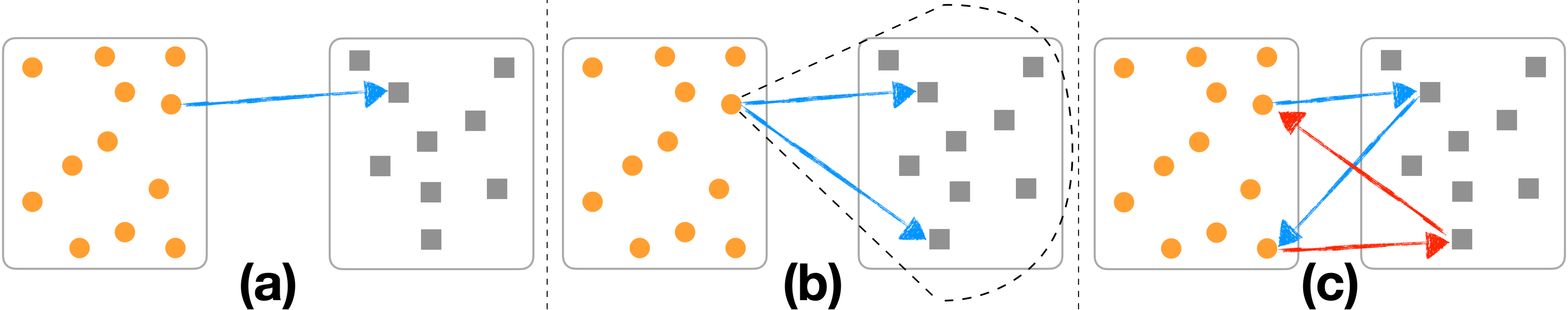}
    \vspace{-7mm}
  \caption{Three types of tracing: (a) \textit{individual} oriented, tracing in a \textit{single} direction; 
  (b) \textit{group} oriented, tracing in a \textit{constrained} direction, 
  and (c) \textit{cluster} oriented, tracing in \textit{reflected} directions. 
  A blue/red arrow indicates tracing direction and dotted lines show a constrained range of tracing.
  }
  ~\label{3tracings}
  \vspace{-2mm}
\end{figure}

Considering the structure of cross-view data relationships (see Section \ref{cvrelation}), there are three primary types of tracing (see Figure \ref{3tracings}): 1) \textit{individual} oriented tracing, 2) \textit{group} oriented tracing, and 3) \textit{cluster} oriented tracing.
These correspond to the first three levels of cross-view data relationships outlined in Table \ref{4levels}.

\textbf{Individual oriented tracing} (\textit{T1}) follows a single related visual element in a \textit{single} direction, reflecting one-to-one relationships.
\textbf{Group oriented tracing} (\textit{T2}) tracks relationships between a single element and multiple related elements within a bounded range, resulting in \textit{constrained} directional tracing.
Thus, it involves a group-level relationship (a one-to-many relationship).
\textbf{Cluster oriented tracing} (\textit{T3}) examines relationships between two groups of elements, involving \textit{reflected} directional tracing.
This reveals a bi-group level relationship (a many-to-many relationship).
While more complex relationships (multi-group level) exist, tracing them requires progressive analysis through multiple group or cluster-oriented traces, as they involve chains of information across multiple views.
Therefore, this study focuses on the three primary types of tracing mentioned above.

\subsection{Design Analysis: Trade-off \& Consideration}
\label{trade-offs}

\begin{figure}[tb]
  \centering
  \includegraphics[width=\columnwidth]{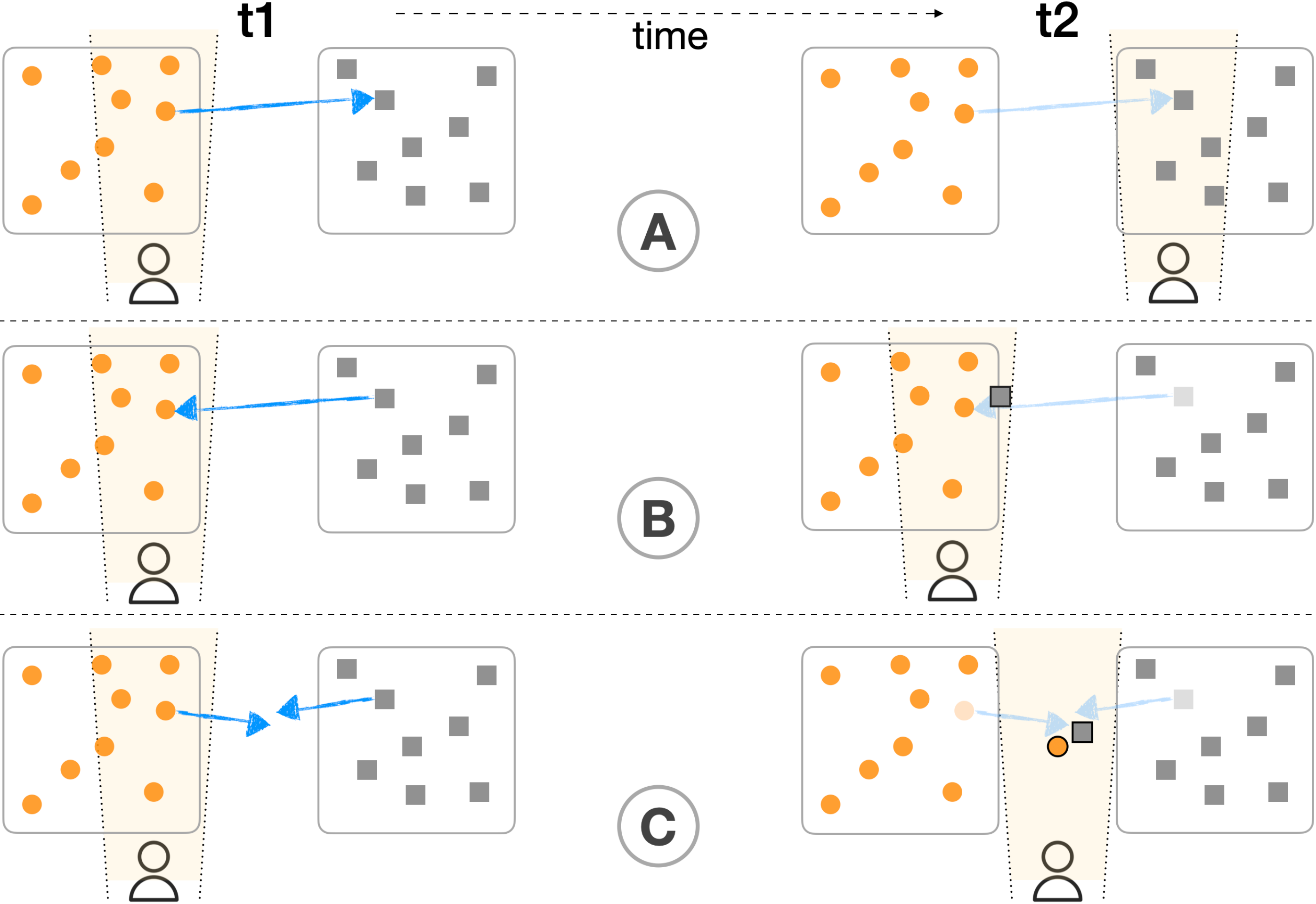}
    \vspace{-7mm}
  \caption{Designs for supporting tracing:
  context \textit{switching} (A): user focus shifts from one view to another,
  context \textit{enriching} (B): moving related visual elements from another view to the current one, and
  context \textit{separating} (C): placing related visual elements outside original views and near each other. 
  A trapezoid shows a user's focus at a time.
  A blue arrow reveals a moving direction.
  }
  ~\label{3designs}
  \vspace{-2mm}
\end{figure}

Considering the change of a user's focus, as shown in Figure \ref{3designs}, there are three possible design strategies for supporting tracing:
1) \textit{context switching}, 2) \textit{context enriching}, and 3) \textit{context separating}.

The \textbf{context switching} (S1) design strategy focuses on the necessity for users to transition their focus across views (Figure \ref{3designs}(A)).
It acknowledges human cognition's limited capacity and the spatial separation of views, assuming users can only focus on specific elements within a single view at a time. 
While this approach preserves each view's original organization, it has notable drawbacks.
The spatial separation of related elements and constant context switching increases cognitive load \cite{convertino2003exploring}.
Users need clear visual cues to track related elements across views, especially when views are far apart.
Without such cues, users can be confused about which visual elements in the current view correspond to those previously checked.

The \textbf{context enriching} (S2) design aims to keep a user's analysis focus within a single view by reducing the need to switch the focus between views (Figure \ref{3designs}(B)).
This is achieved by moving 
related visual elements from other views into the current working view.
This strategy enriches the visual context of the analysis, allowing users to trace cross-view data relationships 
without constant focus shifts, thereby reducing cognitive effort.
However, this introduces new challenges. 
Moving elements between views requires reorganizing them, which can impact their structure. 
Users have to work on views in \textit{transformative} forms, where visual elements in views are no longer confined within the traditional boundaries (i.e., visible borders) of a view. 
Instead, elements can be flexibly moved based on cross-view data relationships. 
This flexibility blurs boundaries, potentially complicating perception and leading to visual clutter. 

The \textbf{context separating} (S3) design challenges the existing organizations of MVs by moving all related visual elements out of their original views (Figure \ref{3designs}(C)).
It separates related visual elements from unrelated ones, effectively dividing a display space into two parts: 
1) original views and 2) relationships among views.
The former are the views as initially designed.
The latter are areas where related elements are extracted into new types of views (e.g., relationship views).
Such a division allows users to easily identify related visual elements, as they sit together and segregated from unrelated information, minimizing interference and visual clutter.
Positioning related elements outside views can bring them into closer proximity than when dispersed across different views, minimizing the effort required for tracing due to reduced physical movement.
However, this separation poses new challenges. Relocating visual elements strips them of their initial visual context, which might complicate users' comprehension of cross-view data relationships.

Supporting tracing with these design approaches involves either redirecting a user's focus between views or relocating visual elements outside of their original views. 
The first strategy maintains the structure of views at the cost of requiring users to exert effort in transitioning between them. 
The second reduces user effort by minimizing the need to switch between views but disrupts the organization of views. 
The third allows users to easily identify connections without interference by segregating related elements from unrelated ones but strips elements of their original visual context. 
These compromises prompt a reevaluation of the standard organization of MVs.

Based on our analysis of user tasks, types of tracing, and design strategies with trade-offs, we have identified four considerations that the design of \name\ attempts to follow.

\textbf{C1: Providing usable visual guidance to support transitioning between views}. Transitioning between views demands significant cognitive effort \cite{convertino2003exploring}, which commonly occurs while using MVs, so it is vital to offer usable visual guidance to help users locate the target segment when moving from one view to another. 
Specifically, such visual guidance should direct a user's attention to three key aspects of cross-view transitions involved in tracing: \textit{direction}, \textit{path}, and \textit{destination}, corresponding to three main questions for tracing: which view(s) to navigate to, how to get there, and where to stop within the target view(s). 
(a) Showing the tracing \textit{direction} helps users understand which view(s) they need to transition to.
(b) Providing transition \textit{paths} directs users' attention during a transition process and helps to maintain focus.
(c) Highlighting related visual elements within the target view(s) guides users to the \textit{destination} elements they need to trace, effectively concluding the transition process.

\textbf{C2: Scaffolding multi-directional tracing}.
As related visual elements are spatially scattered, it is important to offer techniques to support users in tracing multiple directions.
Given the impact of selective attention \cite{treisman1969strategies}, it is hard for users to trace several directions simultaneously, especially in divergent or opposite directions.
While it is possible to turn this into a sequential process in which each step only involves tracing in one direction (\textbf{T1}), transitioning from a one-directional tracing to another creates challenges (e.g., which direction to pursue next and how users can ascertain completion of tracing in all directions).
As multi-directional tracing (\textbf{T2}, \textbf{T3}) has been commonly involved in real-world analyses, developing techniques to facilitating this is imperative.

\textbf{C3: Comprehensive tracing of relationships}.
To foster a transparent understanding of intricate interrelations among elements, it is crucial to allow users to navigate through all relationships associated with a visual element.
In complex datasets, elements often have multiple connections spanning across elements in different views. 
Users should be able to confidently explore these connections.
By supporting comprehensive tracing, users can efficiently trace individual elements (\textbf{T1}), follow connections between an element and a group (\textbf{T2}), and understand how groups relate to each other (\textbf{T3}).
Without the ability to trace them, users may overlook critical connections, leading to incomplete analyses or incorrect conclusions.

\textbf{C4: Dynamic transition via tracing}.
Dynamic transitions can be used to reduce visual complexity while a user is engaged in tracing an element.
Factors influencing visual complexity include the number of elements, visual links, and views. 
Reducing visual complexity facilitates tracing and helps users effectively explore significant connections.
Additionally, enhancing the salience of visual links supports users in tracing connections, allowing them to maintain focus and differentiate desired links.
Informing users of their progress during tracing can improve their experience, aligning with usability heuristics (i.e., visibility of system status and recognition rather than recall) discussed by Nielsen and Molich \cite{nielsen1990heuristic}. 
Awareness of position and progress clarifies task completion, helping users see what has been explored and what remains.
This feedback aids in resuming interrupted tasks and reduces cognitive load, especially when tracing spatially distant elements across multiple views.

%% file: tex/4Technique.tex
\section{The \name\ Technique}
\label{sec-tech}

\begin{figure}[tb]
  \centering
  \includegraphics[width=\columnwidth]{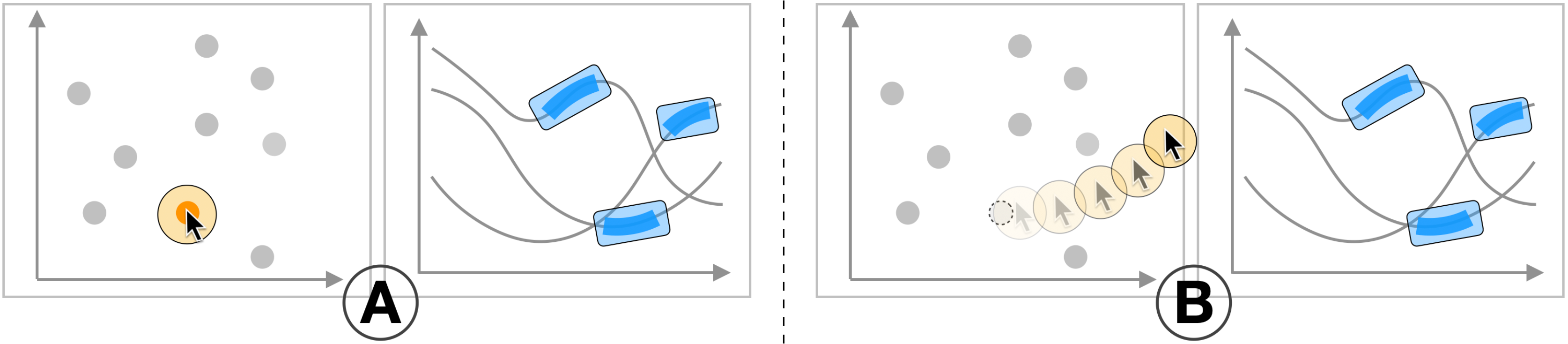}
    \vspace{-7mm}
  \caption{Examples of externalizing a user's focus in \name.
  (A): \name\ creates a copy of each visual element involved in a cross-view data relationship and overlays them on existing views.
  (B): \name\ enables automatically moving a focus marker as a user moves a mouse pointer from its previously focused visual element.
  Orange circles and blue rectangles show the focus marker and copy of related visual elements in \name, respectively.
  }
  ~\label{externalization}
  \vspace{-3mm}
\end{figure}

The \name\ technique highligths a design concept of \textit{interactive focus transition} to support cross-view data relationship tracing, which aims to make a user's focus visible and manipulable. 
It has three components: 1) externalizing a user's focus during tracing, 2) manipulating a user's focus to facilitate tracing, and 3) dynamic transition of a user's focus.
They are applied on top of identified cross-view data relationships to support users in tracing them, so \name\ requires that cross-view data relationships are computed in advance.
To compute the relationships, \name\ follows prior work \cite{sun2021sightbi}, which uses \textit{biclustering} to compute relationships between visual elements of pairwise views and chaining of biclusters (i.e., results of biclustering algorithms) to identify relationships among visual elements of three or more views.
The computed cross-view data relationships serve as the foundation of \name, enabling it to provide visual cues and interactions that help users trace connections.
They can be shown in various means, such as visual links connecting related elements, highlights that emphasize relationships, and dedicated relationship views that display visual elements together \cite{sun2021sightbi}.

\subsection{Externalizing User Focus}
\label{sec-externalization}

To make a user's focus explicit, \name\ introduces visual overlays, called \textbf{focus markers}, to show the user's current focus.
Users can enable or disable a focus marker by right-clicking on a visual element. 
When enabled, a semi-transparent, slightly larger copy shows over the element (orange circles in Figure \ref{externalization}), distinguishing it from the original visual element without obscuring essential details.
By interacting with multiple visual elements, users can create several focus markers simultaneously, allowing the tracing of bi-group level relationships (\textbf{T3}).
Moreover, based on computed cross-view data relationships, \name\ enables users to generate copies of related visual elements in other views (blue rectangles in Figure \ref{externalization}).
These copies can also be enabled or disabled via a right-click menu on each focused visual element.
They share visual cues with focus markers and represent \textit{related elements}. 
This helps users recognize copies of visual elements, differentiate visual elements from their copies, and enable direct manipulations on focus markers.

\begin{figure*}[tb]
  \centering
  \includegraphics[width=\textwidth]{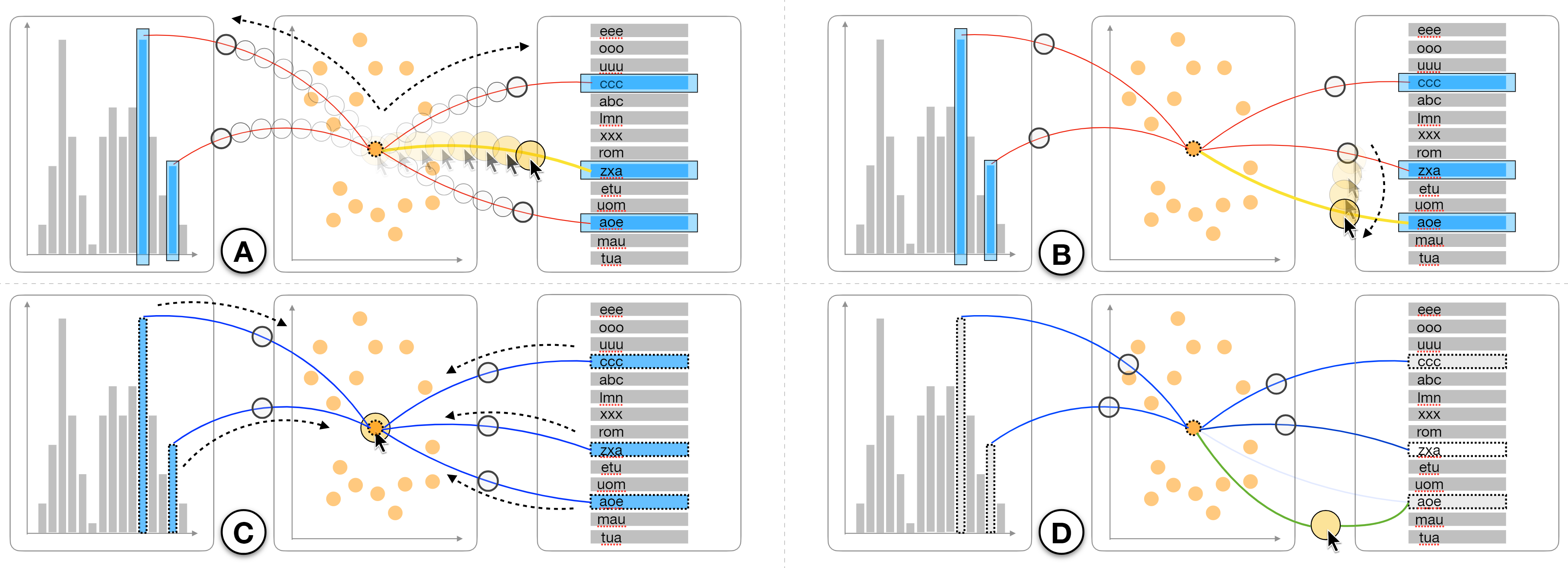}
    \vspace{-7mm}
  \caption{Examples of manipulating externalized user focus in \name. 
  (A): When a user moves the focus marker (i.e., a copy of a visual element in a scatterplot) along a visual link, supportive foci are displayed and automatically move along other visual links from the previously focused visual element.
  (B): A user moves the focus marker from one visual link to another.
  (C): From a currently focused visual element, a user calls for its related visual elements from other views, which shifts their copies along corresponding visual links.
  (D): A user interacts with a focus marker to manually manage links, positioning them as desired.
  Note: black dotted arrows reveal the direction of transitions, and the green link indicates a manually managed link.
  }
  ~\label{move}
\end{figure*}

While the current implementation uses a semi-transparent, scaled copy for the focus marker, the \name\ technique is designed to be extensible, allowing its visual encoding to be customized for better visibility in complex scenarios, as noted in our user study. 
Instead of a direct copy, the marker could use distinct shapes (e.g., star, triangle, outlined circle), have static or density-aware sizing, or be enlarged relative to the original. 
Additional enhancements include high-contrast or saturated colors, stronger outlines or shadows, subtle animations (e.g., pulsing, blinking), or alternative fill styles like hatching. 
These adaptations can be tailored to view characteristics, information density, or user preference to ensure the marker remains easily distinguishable during tracing.

In \name, visual salience can propagate across multiple views based on computed cross-view data relationships, showing both a user's current and future focus.
For example, in Figure \ref{externalization}, orange circles correspond to a user's current focus (i.e., where a tracing starts).
Blue rectangles indicate the user's future focus (i.e., where the tracing will end).
\name\ uses the visual salience in other view(s) to offer users guidance on tracing direction and destination (\textbf{C1-a, c}).   
After creating focus markers and copies, \name\ allows users to move these markers manually or automatically.
For \textit{automatic movement}, a focus marker moves by following a user's mouse pointer (Figure \ref{externalization} (B)), externalizing the transition of the user's focus by transforming it from an internal cognitive process into visible on-screen elements.
For \textit{manual movement}, a user can drag and move the marker flexibly, which is discussed with more detail in Section \ref{manipulation}.

\subsection{Manipulating User Focus}
\label{manipulation}

\begin{figure}[tb]
  \centering
  \includegraphics[width=\columnwidth]{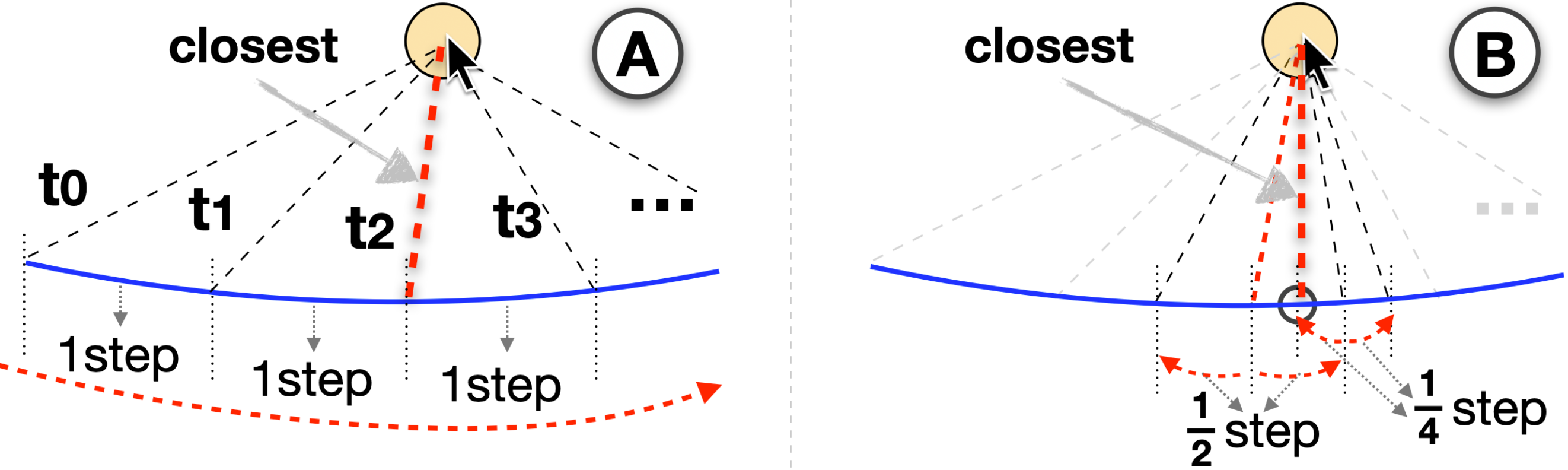}
    \vspace{-7mm}
  \caption{\name's two procedures to find the closest point on a visual link by the position of a user's focus: (A) a linear search process, and (B) an iterative bidirectional search process.
  }
  ~\label{fig-algo}
  \vspace{-3mm}
\end{figure}

In \name, focus markers are considered the \textit{first-level objects} for user interaction.
Users can manipulate these markers by moving them to attract related visual elements from other views (Figure \ref{move}).
Besides the automatic movement discussed in Section \ref{sec-externalization}, users can manually move a focus marker by dragging it.
To guide users in moving their focus marker, \name\ uses \textit{visual links} (Figure \ref{move} (A), (B)) from a visual element to its related elements, explicitly showing possible paths for users to perform tracing (\textbf{C1-b}). 
Users can move focus markers along these links (Figure \ref{move} (A)).
\name\ supports \textit{cross-link transitions} with the focus marker (Figure \ref{move} (B)).
During the movement of a marker by a user, \name\ detects the closest point on any visual link connected to the current visual element and adjusts the marker's position accordingly.
To find the closest point on a visual link, \name\ uses an algorithm with a linear \cite{cormen2022introduction} and iterative bidirectional search process \cite{russell2020artificial}. 
Figure \ref{fig-algo} shows the two procedures in this algorithm (see Algorithm 1).
This algorithm is also used for automatic movement, ensuring that the focus marker "sticks" to the nearest visual link relative to the user’s mouse pointer. 

Along with the transition of a user's focus marker, \name\ creates \textbf{supportive foci} (white circles in Figure \ref{move} (A) and (B)) that move along all other visual links associated with the element being traced.
These supportive foci can be enabled or disabled by left-clicking on the marker and are designed to assist in multi-directional tracing (\textbf{C2}) and facilitate comprehensive tracing of relationships (\textbf{C3}).
This supports the context switching (\textbf{S3}) strategy, enabling users to shift their focus between related visual elements across multiple views.
Supportive foci move in sync with the user's focus marker, with \name\ calculating their movement in real time based on the proportion of the marker's transition along its visual link.
These supportive foci help users identify visual links that share the same cross-view data relationship as the focus marker. 
This reduces the effort required to locate relevant links, particularly in cases where many unrelated visual links are present. 
By acting as visual anchors, the supportive foci guide a user's attention to relevant connections.
Research shows that humans better distinguish shapes (e.g., circles and lines) than color differences \cite{munzner2014visualization}.
\name\ leverages this by using circles for supportive foci and lines for visual links.
Moreover, when a user's focus marker transitions from one visual link to another, the supportive foci on the new link switch as well (Figure \ref{move} (B)).

\begin{algorithm}[tb]
\scriptsize
\caption{ClosestPoint(path, point)}
    \SetKwInOut{Input}{Input}
    \SetKwInOut{Output}{Output}

\SetKwData{pathNode}{path}
\SetKwData{selPoint}{closestPoint}
\SetKwData{point}{point}
\SetKwData{best}{bestPoint}
\SetKwData{scan}{scanPoint}
\SetKwData{scanLength}{scanLen}
\SetKwData{scanDistance}{scanDist}
\SetKwData{bestLength}{flagLen}
\SetKwData{bestDistance}{closestDist}
\SetKwData{precision}{step}
\SetKwData{pathLength}{pathLen}
\SetKwData{beforeLength}{left}
\SetKwData{afterLength}{right}
\SetKwData{before}{leftPoint}
\SetKwData{after}{rightPoint}
\SetKwData{beforeDistance}{leftDist}
\SetKwData{afterDistance}{rightDist}

\SetKwFunction{DistanceTwo}{EuclideanDist}
\SetKwFunction{getTotalLength}{getTotalLength}
\SetKwFunction{getPointAtLength}{getPointAtLength}
\SetKwRepeat{Do}{do}{while}

\SetCommentSty{mycommfont}

    \Input{\pathNode, a path object; \point, a point object}
    \Output{\selPoint, the closest point object on a path}

\selPoint, \bestLength, \bestDistance $\leftarrow$ none, 0, $\infty$\;
\precision, \pathLength, \scanLength $\leftarrow$ 8, \pathNode.\getTotalLength(), 0\;

\While(\tcc*[f]{linear search}){\scanLength $<$ \pathLength}{
    \scan $\leftarrow$ \pathNode.\getPointAtLength(\scanLength);
    
	\scanDistance $\leftarrow$ \DistanceTwo(\scan, \point);
    
    \If{\scanDistance $<$ \bestDistance}{
        \selPoint, \bestLength, \bestDistance $\leftarrow$ \scan, \scanLength, \scanDistance\;
    }
    \scanLength $\leftarrow$ \scanLength + \precision;
}
\While(\tcc*[f]{bidirectional search}){\precision $>$ 0.5}{
	\precision $\leftarrow$ \precision $/$ 2;
	
    \beforeLength, \afterLength $\leftarrow$ \bestLength\ - \precision, \bestLength + \precision\;
    \before, \after $\leftarrow$ \pathNode.\getPointAtLength(\beforeLength), \pathNode.\getPointAtLength(\afterLength)\;
    \beforeDistance, \afterDistance $\leftarrow$ \DistanceTwo(\before, \point), \DistanceTwo(\after, \point)\;

    \If{\beforeDistance $<$ \bestDistance}{
        \selPoint, \bestLength, \bestDistance $\leftarrow$ \before, \beforeLength, \beforeDistance\;
    }
    \ElseIf{\afterDistance $<$ \bestDistance}{
        \selPoint, \bestLength, \bestDistance $\leftarrow$ \after, \afterLength, \afterDistance\;
    }
}
\KwRet{\selPoint}\;
\end{algorithm}

\name\ also allows the transition of related visual elements, 
using a \textit{magnet and dust} visual metaphor \cite{soo2005dust}.
From a focus marker (serving as the magnet, \textbf{C1-a}), the copy of its related visual elements (considered as the dust) can be attracted from other views (Figure \ref{move} (C)).
These copies move along the visual links that connect related elements to the focus marker (\textbf{C1-b}).
Users can enable such transitions with a right-click menu on the element.
\name\ offers two options in the right-click menu to control where the transitions stop: (\textbf{C1-c}): 1) at the border of the view where the focus marker is located (Figure \ref{move} (C)), or 2) near the focus marker itself.
The former aligns with the context-enriching design (\textbf{S2}) as user focus is maintained in a view.
The latter supports the context-separating design (\textbf{S3}) when the focus marker is positioned away from the views.
Users can hover over each moved copy and supportive foci to show information (e.g., an entity label) of its corresponding visual element in a popup.
Moreover, when hovering over a copy, its corresponding visual mark in the original view is highlighted.

Moreover, \name\ introduces a \textit{manual link management} feature that allows users to interact with visual links using focus markers to drag and reposition them (Figure \ref{move} (D)).
Users can manage visual links manually, placing them anywhere within the display space.
This is toggled via a right-click menu on the focus marker.
Once activated, the focus marker moves along a user's mouse cursor dragging the visual link it was connected to at the time of activation. 
This feature allows users to store specific connections they may wish to explore later, effectively bookmarking relationships for future analysis.
It also facilitates a more comprehensive tracing of relationships (\textbf{C3}) between two elements (\textbf{T1}) by providing users the flexibility to examine connections at their own pace.
\name\ also enables the link management of multiple visual links. 
Users can make non-pinned links, either semi-transparent or hidden, via a right-click menu, allowing them to focus exclusively on the pinned connections.
This enhances the flexibility of tracing, supporting both group-level tracing (\textbf{T2}) and bi-group level tracing (\textbf{T3}).

\subsection{Dynamic Transition of User Focus}

As a focus marker moves along a visual link, \name\ uses dynamic transitions to support comprehensive relationship tracing (\textbf{C4}).
\name\ uses color encoding to indicate which link a user is focusing on.
The visual link currently being traversed is highlighted in yellow, marking it as the active link.
Meanwhile, related links connected to the same visual element, but not currently followed, are displayed in red, and all other unrelated links are colored blue.
This color scheme allows users to distinguish among the active link, 
the related links available for further exploration, 
and the unrelated links. 

\name\ uses \textit{dynamic transparency control}, allowing users to adjust the visibility of unrelated links as they transition a focus marker. 
Users can choose to reduce the transparency of all unrelated links to focus solely on connections tied to the selected element or fade all links except the one being traced.
The transparency adjusts dynamically with the transition progress.
For instance, when a user reaches the midpoint of a link, unrelated links become 50\% transparent, with opacity decreasing as the user advances.
This progressive adjustment reduces visual complexity while preserving the link structure.
Users can restore visibility by reversing transitions or revisiting previous paths.
These provide users with more control over the display space, allowing for an in-depth exploration while reducing cognitive load.
By highlighting relevant connections and subduing unrelated links, \name\ enhances user confidence in exploring relationships (\textbf{C3}).
A challenge in designing dynamic transitions is maintaining effective visual emphasis in complex multi-view environments. 
We address this using a high-contrast color scheme and dynamic transparency, which highlights the active path and related options while fading unrelated links into the background. 
This reduces clutter and supports user focus, aligning with established foreground-background highlighting strategies in visualization \cite{hall2016formalizing}.

\begin{figure}[tb]
  \centering
  \includegraphics[width=\columnwidth]{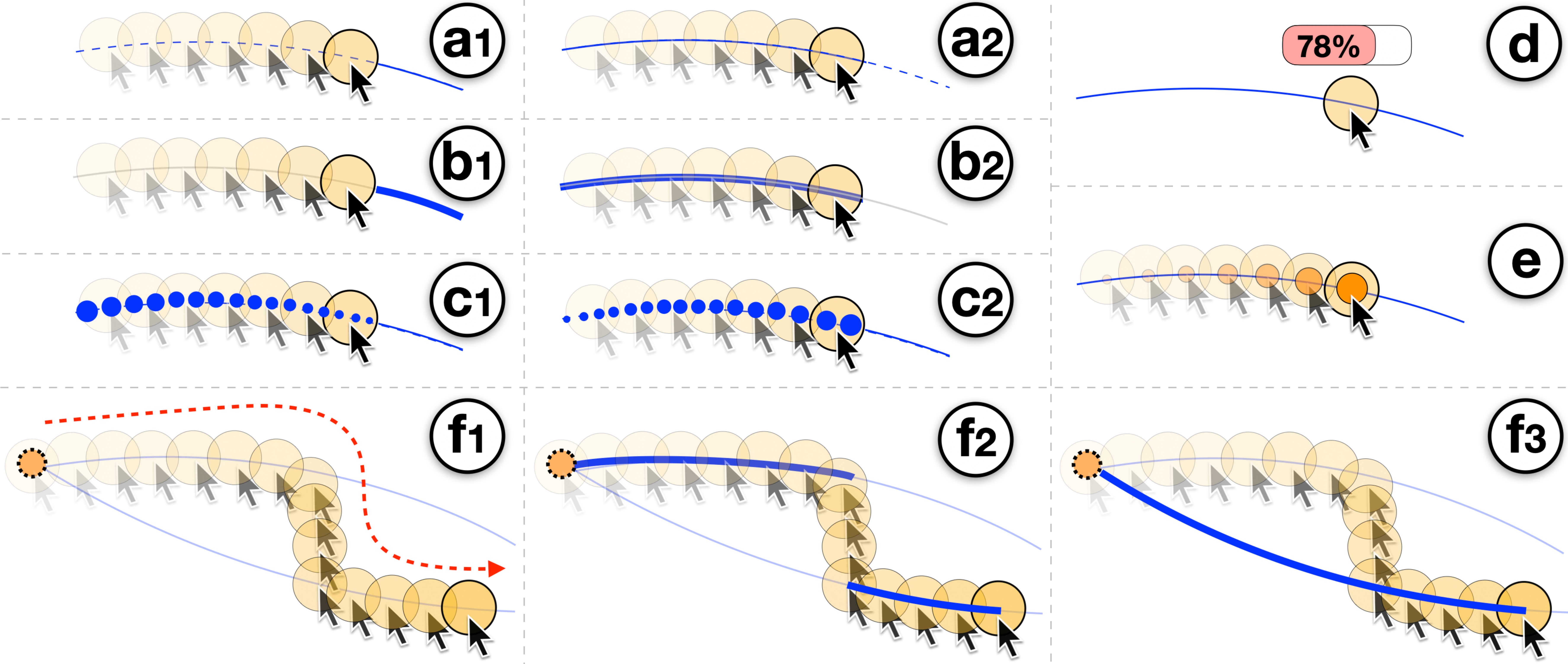}
    \vspace{-7mm}
  \caption{Design alternatives to show a transition progress of a user's focus marker.
  (a2), (b2), and (c2) are encodings focusing on the path that a focus marker has passed.
  (a1), (b1), and (c1) show encodings emphasizing the remaining path.
  (d): Using a progress bar.
  (e): Using a proportional filling to show the progress (i.e., the more a focus marker has transitioned, the more filled area is inside the focus).
  For cases with shifts across links, three designs are considered: (f1) no path highlight, (f2) highlighting the path on each link that a focus marker has passed, and (f3) highlight on the current visual link only (to the most recent position of the focus marker), along which a focus marker is moving.
  }
  ~\label{progress}
  \vspace{-3mm}
\end{figure}

As a focus marker moves, we explored visual aids to display the traced path, tracing direction, and transition progress (\textbf{C1}). 
Figure \ref{progress} shows our design alternatives.
During the transition, the visual link is treated as two segments: the path traversed by the focus marker and the remaining part.
Visual distinctions mark these segments: using dashed versus solid lines, varied line thickness, or dots of different sizes.
While these designs indicate the traced path and direction, they do not effectively show the transition progress.

Focus markers navigate links in two ways: 1) \textit{single-link} transitions along one path (Figure \ref{progress} (a1, b1, c1)), and 2) \textit{multiple-link} transitions between different paths (Figure \ref{progress} (f2)).
For single-link transitions, the traced path, direction, and progress of a focus marker transition can be displayed by either adjusting the visual appearance of the visual link (Figure \ref{progress} (a1), (a2), (b1), (b2), (c1), and (c2)) or the focus marker itself (Figure \ref{progress} (e)), or adding additional marks (Figure \ref{progress} (d)).
One design alternative uses \textit{proportional filling} (Figure \ref{progress} (e)).
It uses a circle within the focus marker that expands during transition.
As a marker's size may limit the visibility of such changes, 
another alternative employs a \textit{progress bar} design (Figure \ref{progress} (d)), where a bar is placed above the focus marker, offering better clarity in tracking movement compared to modifications of the link or marker alone.

For transitions with shifts across links, we explored three designs: 1) no path highlight, 2) highlighting all traversed links, and 3) highlighting only the active link up to a focus marker's position (Figure \ref{progress} (f1), (f2), and (f3), respectively).
The first design shows only the marker position and gets ambiguous when multiple links intersect, making path tracking hard. 
The second design traces the complete transition path but creates visual clutter with numerous highlighted segments across multiple links.
To balance clarity and usability, we adopted a third design that highlights only the current visual link from the trace origin to a focus marker's position.
It reveals the transition process while maintaining clarity, even when other neighboring links are present.
Our final design in \name\ highlights the current visual link being transitioned only (Figure \ref{progress} (f3)), accompanied by a progress bar positioned above the focus marker (Figure \ref{progress} (d)).

%% file: tex/5Study.tex
\section{User Study}

\begin{figure}[tb]
  \centering
  \includegraphics[width=\columnwidth]{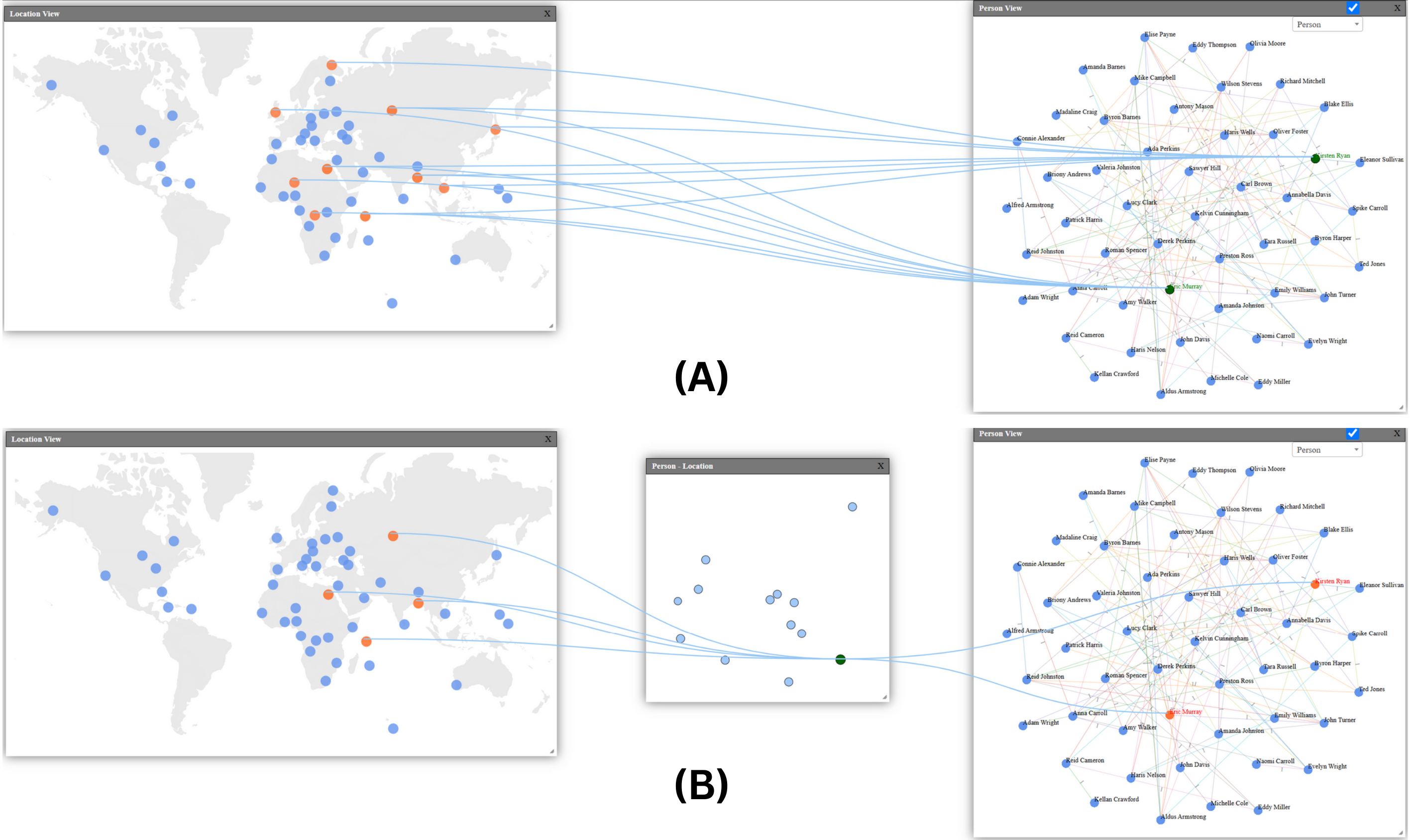}
    \vspace{-7mm}
  \caption{Visualization of the bundling strategy.
  (A): \textit{Without} bundling -- individual visual links connect related entities across views, resulting in visual clutter.
  (B): \textit{With} bundling -- common relationships are aggregated and routed through a bundle element in a relationship view, simplifying the structure and revealing shared group-level connections.
  }
  ~\label{bundling}
  \vspace{-3mm}
\end{figure}

To evaluate the usability of the key design concept, \textit{interactive focus transition}, implemented in \name, we conducted a user study. 
Our study aimed to explore the effectiveness of this design concept for visual analysis involving tracing cross-view data relationships. 
We focused on three questions.
First, would users prefer our interactive techniques over traditional MV techniques for tracing cross-view data relationships and why? 
This helps us understand the comparative appeal of our approach.
Second, how do users apply our interactive techniques in practice? 
Observing their usage patterns offers insights into the intuitive use of our design.
Finally, what do users perceive as the strengths and weaknesses of our techniques? This feedback is key to understanding the benefits and identifying areas for improvement.

\subsection{Participants, Apparatus and Data}

30 graduate students (14 female and 16 male) from several departments in a university, aged 23 to 33 years (mean age = 27),
participated in our study voluntarily.
All had normal or corrected-to-normal vision, no color vision deficiencies, and prior experience using multiple views.
We deployed \name\ on a laptop with a 2.3 GHz Intel Core i7 processor and 16GB of memory, connected to a 49-inch ultrawide monitor with a resolution of 5120 $\times$ 1440 pixels.
Visualizations were shown in Google Chrome (version 116, 64-bit), which fit the screen without requiring scrolling. 
Participants sat approximately 55 cm from the monitor and used a mouse and keyboard for interaction.
All participants provided informed consent prior to the study.

\begin{figure}[tb]
  \centering
  \includegraphics[width=\columnwidth]{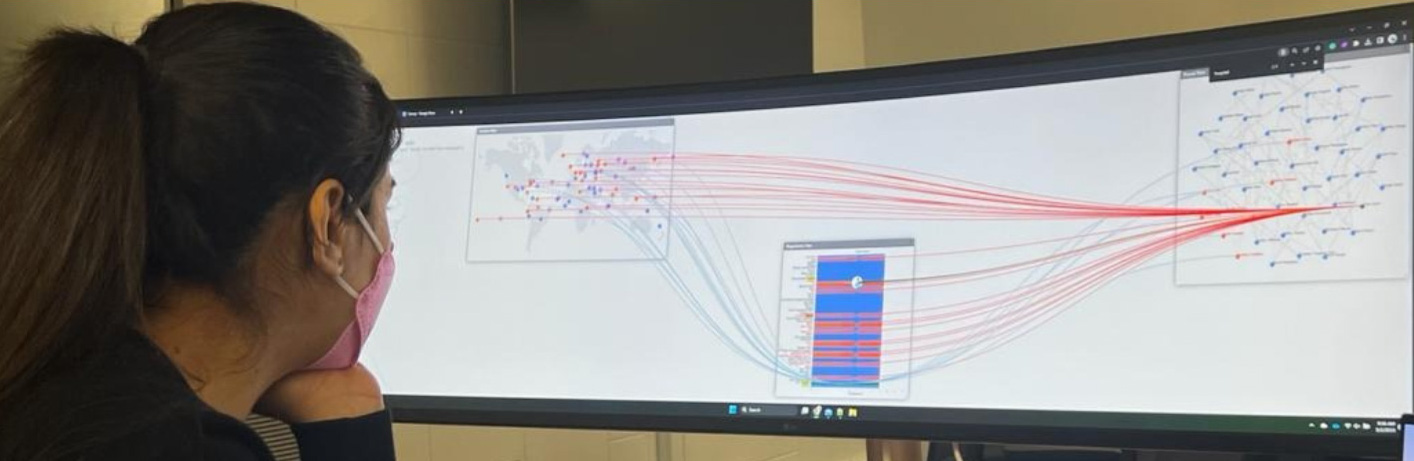}
    \vspace{-7mm}
  \caption{An example of the user study setting 
  }
  ~\label{study-setting}
  \vspace{-3mm}
\end{figure}

We generated eight datasets, each having three types of entities (e.g., person, location, and organization), with 50 entities of each type. 
Each type of entities was shown in a view: the person entities in a network graph, location entities on a map, and organization entities in a bar chart.
We established cross-view data relationships by defining individual-level relationships between entities from different types (e.g., persons associated with locations).
These relationships were set at two levels: by picking 10\% or 20\% of all pairwise entities (i.e., 2500 pairs) from two sets.
Using these individual-level relationships, we applied LCM \cite{uno2004efficient} to compute bi-group relationships between pairs of entity sets, ensuring that each group contained at least two entities.
In summary, each dataset contains 3 different entity sets with a total of 150 unique entities, 250 or 500 individual-level relationships, and 8\textasciitilde16 bi-group relationships.
Figure \ref{study-setting} shows an example of the user study setting.

\subsection{Method, Procedure and Data Collection}


This study used a full-factorial, within-subjects design to explore the use of \name\ under varying conditions.
We controlled two key factors: 1) \textbf{data complexity} and 2) \textbf{representation complexity}, as they influence the number of visible cross-view data relationships.
For the former, we tested two levels: \textit{fewer} versus \textit{more} number of relationships in data (i.e., 250 vs. 500 individual-level relationships). 
For the latter, we set two levels: \textit{without} versus \textit{with} bundling visual links (i.e., grouping visual links that are in the same bi-group level relationships).
As the two factors are independent, our study had four experiment conditions.
In each condition, participants used three fixed-position visualizations: a map (left), a bar chart (center), and a network graph (right), each with 50 entities.
In conditions with bundling, relationship views were displayed, grouping visual links that are in the same bi-group relationships as outlined in \cite{sun2021sightbi}.
Specifically, if multiple elements in one view are commonly linked to elements in another view, their connections are aggregated and routed through an intermediate "bundle element" shown in a separate relationship view. 
This approach reduces clutter and emphasizes group-level associations by simplifying redundant links. The bundled connections are visualized explicitly to aid interpretation of shared relationships (Figure~\ref{bundling}).

All of \name's techniques, described in Section \ref{sec-tech}, were available for participants to use in each condition.
Each participant completed four information-foraging tasks (one task per experimental condition).
The tasks were designed to reflect the three major types of tracing scenarios, corresponding to the levels of cross-view data relationships.
An example task is: \textit{"Find the organizations in North America where both the Andrews and Chandler families work."} This task requires participants to trace relationships between specific individuals (persons), locations, and organizations, involving both individual and group-level tracing.
The order of tasks and their mapping to experimental conditions were randomized for each participant to counterbalance any learning effects.

\begin{figure*}[tb]
  \centering
  \includegraphics[width=\textwidth]{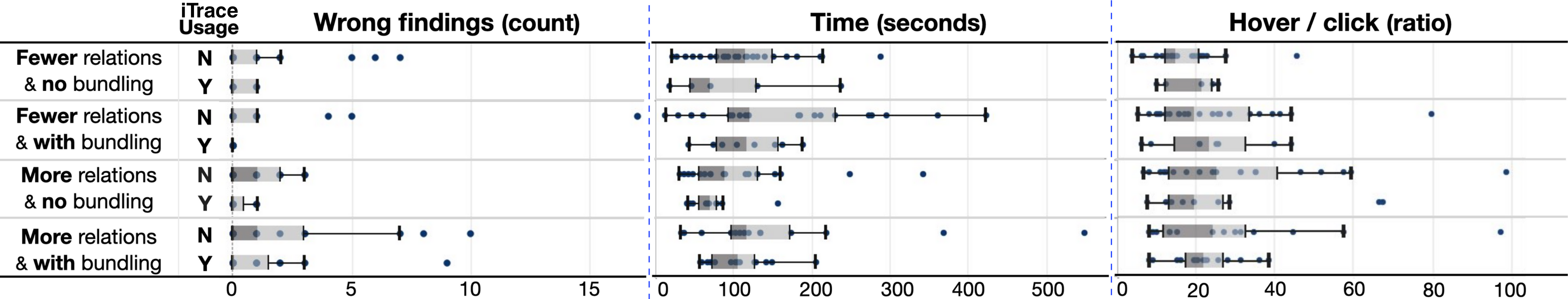}
    \vspace{-7mm}
  \caption{A summary of the number of participants' wrong findings (left), the average time that participants took to get a correct finding (middle), and the ratio between hover and click interactions that participants performed to find answers to the given tasks (right), in each experiment condition.
  }
  ~\label{fig-results}
  \vspace{-1mm}
\end{figure*}

Participants were given 15 minutes per task and could freely use any interactive techniques provided by \name\ based on their preferences. 
After each task, participants reported their findings and could take breaks as needed.
Before the tasks, participants received a tutorial on cross-view data relations and multiple views, followed by a demonstration of \name's features using a separate dataset. 
A training session allowed them to practice identifying relationships across views.
During each task, we collected interaction logs (e.g., timestamps, interaction types, and target elements), screenshots, observation notes, and participants' findings, and conducted interviews to gather their feedback.
Accuracy was measured based on the correctness of participants' reported findings compared to the known relationships in the datasets.

\subsection{Results}
\label{study-results}

\begin{figure}[tb]
  \centering
  \includegraphics[width=\columnwidth]{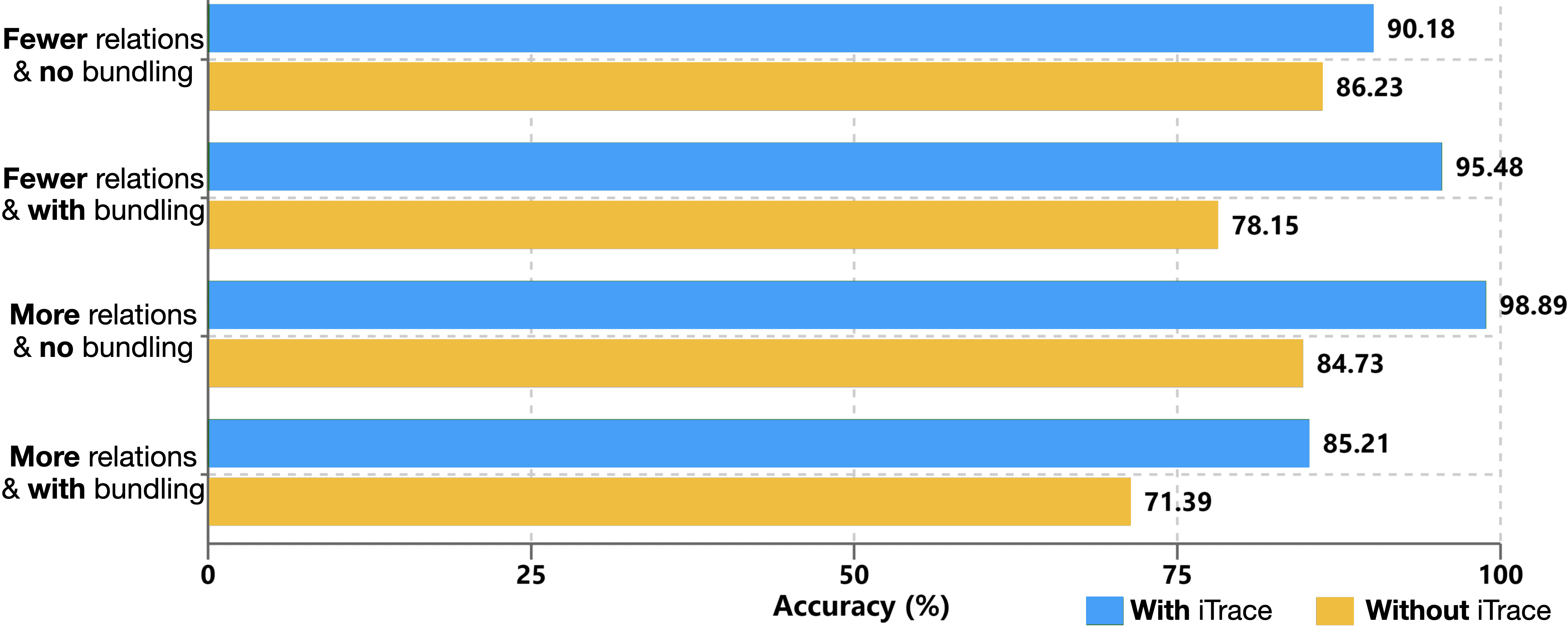}
  \vspace{-7mm}
  \caption{
  A summary of participants' finding accuracy.
  }
  ~\label{fig-ifoc-accuracy}
  \vspace{-3mm}
\end{figure}

We observed that the number of participants using \name\ varied across conditions, and their task performance differed based on if \name\ was used (Figure \ref{fig-results}).
This suggests that the decision to use \name\ may be influenced by the perceived complexity of multiple views, affecting strategic interactions and leading to varied performance.
Despite these variations, participants consistently got higher accuracy when using \name\ (Figure \ref{fig-ifoc-accuracy}).

With the same visual link settings (i.e., with/without bundling), more participants chose to use \name\ when there were more cross-view data relationships.
The increase in relationships led to more visual links.
Participants may perceive this as visually overwhelming.
For example, in conditions with more relationships, hovering over an entity would show a larger number of visual links than those with fewer ones. 
This increased visual complexity likely made it harder for participants to trace cross-view data relations, explaining the higher usage of \name\ in these scenarios.
Moreover, more participants used \name\ when bundling was used than when it was not.
Bundling reduced the number of visual links, but it introduced relationship views. 
While reducing visual links outweighed the increase in views, these new views added more visual elements (as each bundle was shown by an element), and visual links were re-routed through the bundles, making them longer.
It made the representation more complex, encouraging more participants to rely on \name\ for assistance.

Regarding analysis performance, participants who used \name made fewer errors and took less time to find a correct answer than those without using \name\ (Figure~\ref{fig-results} (left) and (middle)).
This was consistent in all conditions.
Even when the number of relationships or bundling settings changed, leading to variations in error rates and task completion time, participants using \name\ performed better overall.
On average, participants using \name\ had an error rate of 0.5, whereas those not using \name\ had an error rate of 1.7. 
A Mann-Whitney U test indicated this difference was statistically significant ($U = 180$, $p = 0.02$). 
Regarding task completion time, participants using \name\ took an average of 9.0 minutes, compared to 12.3 minutes for those not using it.
An independent samples t-test confirmed this difference was significant ($t(118) = 2.85$, $p = 0.005$). 

The ratio of hover-to-click interactions was lower when \name\ was used than when it was not.
Both hover and click interactions highlighted a visual element and its connected visual links. 
The highlights disappeared after hovering, while they remained after clicking.
Thus, the hover-to-click ratio reflects how much effort participants put into observing changes in highlighting.
Without \name, participants seemed to rely more on observing such highlight changes to trace cross-view data relations.
When using \name, participants took less effort for this, as shown by the smaller hover-to-click ratio.
This likely resulted from participants shifting their focus from observing highlight changes to actively manipulating visualized foci with \name.
This redistribution of effort aligned with better analysis performance, as participants made fewer errors and took less time to find correct answers.
Thus, \name\ appears to help users adopt more effective strategic interactions during analysis.

Besides these quantitative results and positive feedback on the ``\textit{intuitive way of usage}" (as noted by P6, P7, P9, and P13), participants' feedback highlighted three roles that \name\ played in supporting tracing cross-view data relations: 1) \textbf{scaling up tracing}, 2) \textbf{diversifying tracing purpose}, and 3) \textbf{strengthening confidence}.
\name\  helped participants scale up their tracing capabilities by allowing them to perform multi-directional tracing even in the presence of many visual links.
For example, P10 mentioned ``\textit{It’s really helpful to set up [see and use] copies to find a node connected to different other ones in other views}"; P1 said, ``\textit{[\name] is very useful especially when you have many things selected, and [it is] more useful when there are more connections}"; 
and P17 appreciated, ``\textit{It [\name] is very useful to find out the exact link when there is more data in the view}."
Additionally, \name\ expanded the purpose of tracing beyond just searching for connections. It also supported verification, helping participants confirm if two visual elements were connected.
For example, P14 stated, ``\textit{[\name] really helped to verify the data exactly and check appropriate connections... [\name] made the task easier to verify the connections}"
; and P7 said, ``\textit{[\name] makes it easy to verify and understand the connections between data points}."
Finally, using \name\ made participants more confident in their analyses. 
For instance, P6 said, ``\textit{It helps me keep track of which line I’m looking at, so I'm more confident [of my findings]}", and P16 commented, ``\textit{[\name] is good to give me the confidence to draw a conclusion}."

Despite the observed benefits of using \name, participants' feedback indicated two possible improvements.
First, while copies of visual elements created by \name\ can be helpful for tracing in multiple directions and find related elements quickly, it can sometimes lead to visual clutter.
For example, P16 said, ``\textit{It [\name] introduced more nodes [copies of visual elements] in the visualizations, and made it slightly confusing due to the visual clutter of many nodes on neighboring links}", and similarly P7 mentioned, "\textit{[copies of visual elements] get congested when there are too many connections}."
These suggest that for visual elements with many visual links, it may be more effective to adjust the position of related visual element copies. Instead of placing them directly along the visual links, repositioning them to nearby areas could help reduce overlap and avoid visual clutter.
Second, the focus marker's visual encoding was perceived as insufficiently salient, especially when many visual links existed.
For example, P15 mentioned, ``\textit{When not moving [focus marker], it is hard to differentiate between its line and other lines}." 
P4 said similarly, ``\textit{[After moving], it is confusing [locate where my focus was]}."
This implies that the current encoding of the focus marker lacks the visibility needed for participants to easily track their focus, particularly when there are many highlighted links and supportive foci.

%% file: tex/6Discussion.tex
\section{Discussion and Conclusion}

We present \name, an interactive technique to support tracing cross-view data relations.
It introduces the concept of \textit{interactive focus transition}, which transforms users' focus into on-screen, manipulatable objects.
By showing and allowing direct manipulation of these foci, \name\ facilitates multi-directional tracing across multiple levels of data relationships while reducing visual clutter associated with traditional linking methods. 
It enhances the flexibility and dynamism of MVs, moving beyond conventional, view-constrained layouts and instilling greater confidence in visual analysis.

\subsection{Generalization}

\begin{figure}[tb]
  \centering
  \includegraphics[width=\columnwidth]{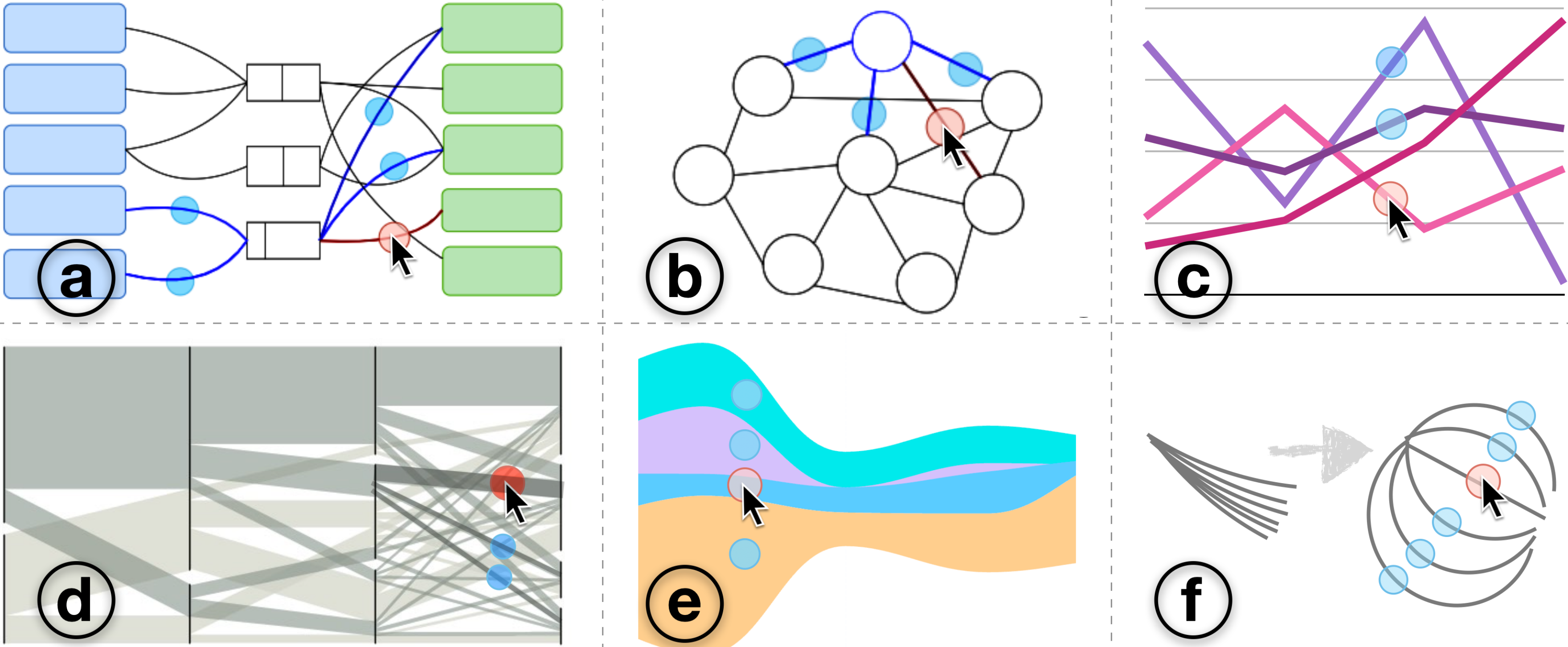}
    \vspace{-7mm}
  \caption{Applying \name\ to other visualizations: (a) lists, (b) graph, (c) line chart, (d) parallel sets, (e) area graph, and (f) pulling nearby lines apart.
  }
  ~\label{fig-generalization}
  \vspace{-3mm}
\end{figure}

The \textit{interactive focus transition} concept extends beyond MV oriented data analysis and can be adapted to a variety of visualizations (Figure \ref{fig-generalization}).
For instance, it can be used on node-link diagrams (e.g., graphs and trees) to check neighboring connections and paths, or to line charts for analyzing temporal trends across one or multiple lines.
For visualizations with variant visual encodings of lines (e.g., ribbons in a Sankey diagram, parallel sets \cite{kosara2006parallel}, or ThemeRiver \cite{havre2002themeriver}), this design concept can also be used.
By assigning forces to supportive foci, the design can improve readability in cases with many overlapping lines by pulling them apart (Figure \ref{fig-generalization} (f)). 
This aligns with existing interactive techniques such as EdgeLens \cite{edgelens}, which employs a lens-based interaction to effectively manage edge congestion in complex graphs.
This directly addresses usability concerns related to visual clutter, as identified in Section~\ref{study-results}.
Compared to traditional visual links and highlights, this design concept makes the user's focus more explicit by turning it into visible, on-screen objects that serve as anchors for attention and facilitate spatial transitions. 
Instead of replacing conventional techniques, it complements them by enhancing their ability to explore related information.
Unlike typical animation techniques that rely on UI widgets (e.g., play and pause buttons), this design emphasizes direct manipulation.
Users can interact with animated objects to control transitions, allowing them to focus on the content rather than the control mechanisms. 
This reduces cognitive load by eliminating the need to switch attention between a visualization and separate control widget(s).


\subsection{Limitations and Future Work}
The implementation of our presented \textit{interactive focus transition} concept in \name\ supports direct manipulations on an individual copy only (e.g., focus marker or one of the supportive foci).
It can be improved by grouping nearby foci moving in similar directions, enhancing \name\ for complex data handling. 
Future studies may explore how to merge and split these foci effectively.
Some techniques, like manual link management and dynamic transitions in \name, introduce an extra level of control that requires further exploration.
While our user study showed the usefulness of \name\ and its possible impact on tracing-based analysis, it was limited (e.g., datasets with a relatively small number of visual elements and views with fixed positions).
Whether our findings would hold with larger datasets or varying distances between views remains uncertain, indicating the need for a more in-depth evaluation.
Furthermore, participants' self-selection to use \name\ during tasks led to unequal group sizes, limiting the ability to definitively attribute performance differences solely to the tool and highlighting the need for future studies with controlled usage conditions.

Additionally, \name\ potentially "breaks" traditional view boundaries by allowing flexible arrangements of visual elements and their copies to support tracing tasks. 
Future studies are needed to explore the potential costs of this approach, such as how users perceive and interact with these ``broken" views.
Performance issues such as lag and choppy interactions during tracing were caused by the real-time rendering demands of dynamic transitions and manipulations involving multiple visual links. 
These limitations can disrupt smooth tracing and negatively affect user experience. Although this study prioritized demonstrating conceptual and practical utility, future implementations could enhance usability by optimizing rendering through caching, incremental loading, or precomputed transitions.